# Determination of the Gate Breakdown Mechanisms in *p*-GaN Gate HEMTs by Multiple-gate-sweep Measurements

Guangnan Zhou, Fanming Zeng, Yang Jiang, Qing Wang, Lingli Jiang, Guangrui (Maggie) Xia, and Hongyu Yu

*Abstract*—In this work, we studied the gate breakdown mechanisms of *p*-GaN gate AlGaN/GaN HEMTs by a novel multiple-gate-sweep-based method. For the first time, three different breakdown mechanisms were observed and identified separately in the same devices: the metal/*p*-GaN junction breakdown, the *p*-GaN/AlGaN/GaN junction breakdown, and the passivation related breakdown. This method is an effective method to determine the breakdown mechanisms. The different BD mechanisms were further confirmed by scanning electron microscopy (SEM). Finally, the temperature dependences of the three BD mechanisms were measured and compared. This analysis method was also employed in the devices with a different passivation material and showed its applicability.

## I. INTRODUCTION

Gallium nitride (GaN) possesses excellent physical properties, such as a high critical electric field and a high saturation velocity [1-3], which is ideal as the semiconductor in a high electron mobility transistors (HEMT) with a low specific ON-resistance ($R_{ON}$), a high breakdown voltage (BV) and a high operation switching frequency. For switching applications, normally-off HEMTs are required to provide adequate safety conditions [4-5]. Among different approaches to realize enhancement mode (e-mode) operation [6-9], a *p*-GaN gate AlGaN/GaN HEMT emerged as a leading solution [10-11].

However, due to the relative low gate BV (usually 10~12 V), the maximum gate operation voltages for *p*-GaN gate HEMTs are usually 5-7 V [10-11]. The small gate voltage swing has imposed significant constraints on the gate driver design. Improving the gate BV remains a critical challenge in *p*-GaN gate HEMTs. However, the gate breakdown (BD) mechanisms are still controversial among the available reliability studies

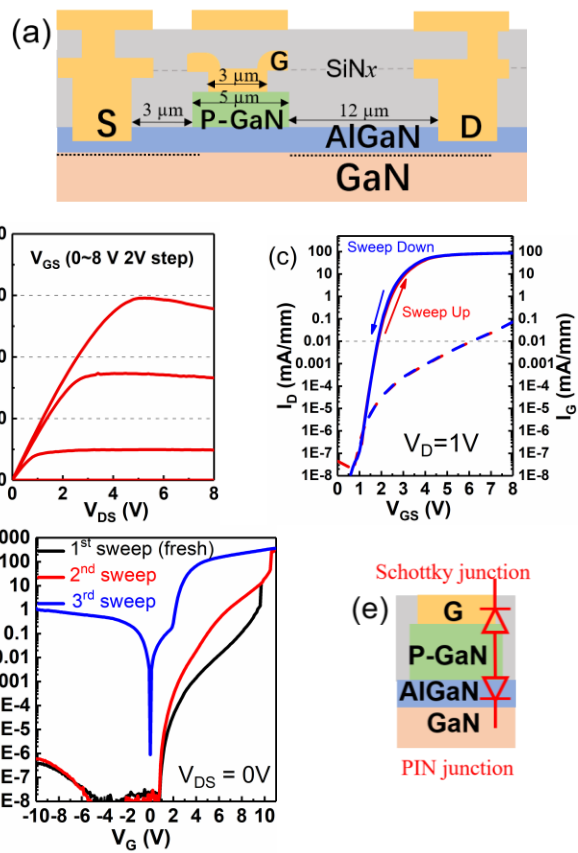

Fig. 1. (a) Schematic of the device structure; (b) output characteristics of the *p*-GaN HEMTs; (c) transfer characteristics of the HEMTs under $V_{DS} = 1$ V; (d) gate leakage characteristics of three gate sweeps with $V_{DS} = 0$ V showing the 1st and the 2nd breakdowns; (e) Schematic cross-sectional view of the gate regions and the equivalent circuit.

[12-22]. Some works have proposed that the gate BD originated from the BD of *p*-GaN/AlGaN/GaN junction (PiN) or the AlGaN barrier layer [12-16]. Meanwhile, others have ascribed

This work was supported by Grant #2019B010128001 and #2019B010142001 from Guangdong Science and Technology Department, Grant #61704004 from Natural Science Foundation of China, Grant #JCYJ20180305180619573 and #JCYJ20170412153356899 from Shenzhen Municipal Council of Science and Innovation. (Corresponding authors: G. Xia and H. Yu.)

G. Zhou is with and the School of Microelectronics, Southern University of Science and Technology (SUSTech) and the Department of Materials Engineering, the Univeriosity of British Columbia (UBC). G. Xia is with the Department of Materials Engineering, UBC. (e-mail: gxia@mail.ubc.ca)

F. Zeng, Y. Jiang, Q. Wang, L. Jiang and H. Yu are with the School of Microelectronics, SUSTech; Engineering Research Center of Integrated Circuits for Next-Generation Communications, Ministry of Education; Shenzhen Institute of Wide-bandgap Semiconductors; and GaN Device Engineering Technology Research Center of Guangdong, SUSTech, 518055 Shenzhen, Guangdong, China. (e-mail: yuhy@sustech.edu.cn)

that to the creation of the percolation path in the *p*-GaN/metal interface [17-19]. Especially, I. Rossetto *et al*. found that the peak electric field across the AlGaN would decrease with the positive gate bias, disapproving the BD of PiN or AlGaN layer [20]. Whether the PiN junction is likely to fail remains indeterminate. A method to determine the BD mechanism is still lacking, which is the motivation of this work.

This paper is to propose and demonstrate a multiple-gate-sweep-based gate BD mechanism analysis approach as an effective and generic method to determine a gate BD mechanism. The multiple-gate-sweep is to induce and decouple the BDs of different device regions. For the first time, three different BD mechanisms have been decoupled and identified from the devices of the same structure: 1) the metal/*p*-GaN junction BD; 2) the PiN junction or AlGaN barrier BD; and 3) the passivation related BD. The PiN junction BD has been directly observed. The BVs of the different BD mechanisms and their dependence on temperature and passivation technology were also investigated.

## II. DEVICE STRUCTURE AND GATE BREAKDOWN

The *p*-GaN gate HEMTs were fabricated on 75 nm *p*-GaN/15 nm $Al_{0.2}Ga_{0.8}N$/0.7 nm AlN/4.5 μm GaN epi-structures grown on Si (111) substrates as shown in Fig. 1(a). The p-GaN layer was doped with Mg to a concentration of $4 \times 10^{19}$ $cm^{-3}$. The fabrication started with p-GaN gate definition by a Cl-based plasma etch followed by $N^+$ ion implantation to isolate the devices. Two SiN layers were deposited as the passivation layers by plasma enhanced chemical vapor deposition (PECVD). A Schottky-type contact was formed between the Ti/Au and the *p*-GaN gate. The devices under test feature a gate width ($W_G$) of 100 μm, a gate length ($L_G$) of 5 μm, a gate-source distance ($L_{GS}$) of 3 μm, and a gate-drain distance ($L_{GD}$) of 12 μm. On-wafer characterization was performed by Keithley 4200 Analyer using the sweep mode of "Normal" measurement speed (0.05V/step, delay time = 1 ms, measure time = 53.59 ms) with a floating substrate. Unless specified, the measurement temperature was 25 °C.

According the circular transimission line model (CTLM) measurements, the sheet resistance of the 2DEG is 610 Ω/square. As shown in Fig. 1(b) and (c), The threshold voltage $V_{TH}$ is extracted to be 1.8 V at $I_D$ of 0.01 mA/mm. A high ON/OFF current ratio of $5 \times 10^8$ and a low on-resistance $R_{ON}$ of 13 Ω*mm have been achieved. At $V_{GS}$ = 8 V, the maximum drain current is extracted to be 290 mA/mm. The device exhibited a breakdown voltage larger than 400V, defined at the criteria of $I_D$ reaching 1 μA/mm.

Fig. 1(e) shows the equivalent circuit of the gate and the regions below, which consists of a metal/*p*-GaN Schottky junction and a *p*-GaN/AlGaN/GaN PiN heterojunction. The Schottky junction limits the $I_G$ when $V_G > 0$ V, while the PiN does that when $V_G < 0$ V. Fig. 1(d) illustrates the gate leakage ($I_G$-$V_G$) characteristics of a typical measurement using three consecutive gate sweeps. The device shows two abrupt $I_G$ increases in the 1st and 2nd sweeps, consistent with very recent studies [21, 22], indicating the existence of at least two different BD mechanisms. Huang *et al.* and He *et al.* have attributed the 1st-step breakdown to Schottky junction failure and 2nd-step breakdown to PiN junction failure. In the following text, we will demonstrate that 2nd-step breakdown should be ascribed to dielectric related failure. Moreover, it's found that the PiN failure won't result in an increase of $I_G$ in the positive bias.

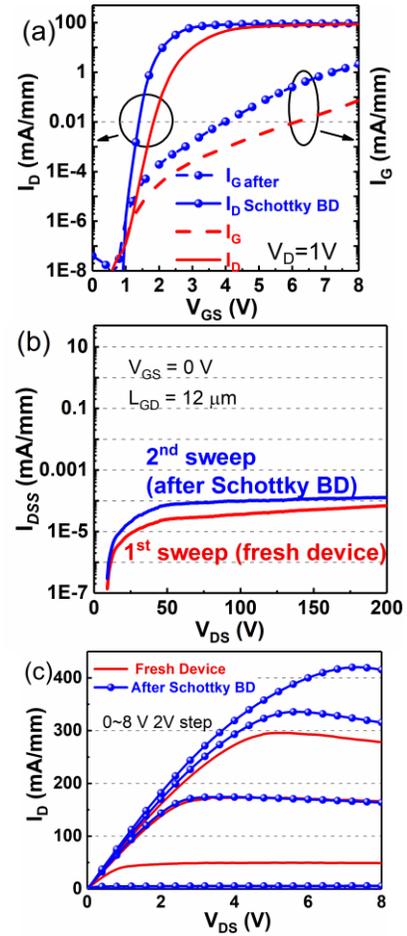

Fig. 2. The device performances before (red) and after (blue) the Schottky junction breakdown: (a) transfer characteristic; (b) OFF-state drain leakage; (c) out characteristic.

## III. RESULTS AND DISCUSSION

### A. The Metal/p-GaN junction BD in the 1st sweep

In Fig. 1, in the first sweep, $I_G$ increases abruptly when $V_G$ reaches 9.2 V, suggesting a hard BD of the gate. However, in the 2nd sweep, the reverse $I_G$ has negligible change compared to the 1st sweep, indicating that the PiN junction remains functional. The transfer characteristics of the HEMTs before and after the 1st BD are shown in Fig. 2(a). The device can still be turned OFF/ON with a high $I_{ON}/I_{OFF}$ ratio. The "gate control" of the channel is preserved. Besides, the device shows a lower threshold voltage, a lower subthreshold swing and a higher drain current ($I_D$) after 1st-step breakdown after the 1st BD. These features indicate that the 1st BD should be attributed to the metal/*p*-GaN Schottky junction failure. The degradation has converted the Schottky junction to an ohmic-like gate. Fig. 2(b) compares the OFF-state drain leakage ($I_{DSS}$) before and after the Schottky BD, showing that the OFF-state leakage blocking capability of the gate stack is maintained.

Although a HEMT cannot function as a normal switch after

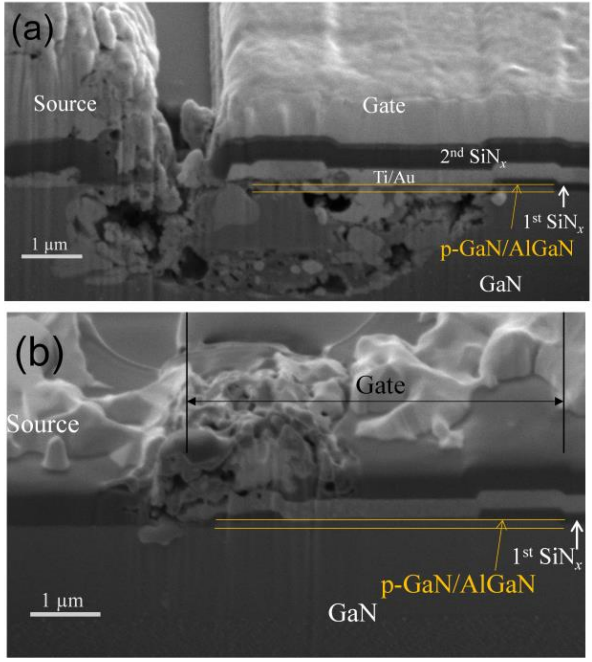

Fig. 6. Cross-section view of BD damages by SEM: (a) damages mainly caused by a PiN junction failure; and (b) damages mainly caused by a passivation related failure. The two parallel lines indicate the top and the bottom surfaces of the AlGaN layer.

responsible for the passivation related BD.

the 1$^{st}$ BD, the following BDs in the subsequent sweeps of the HEMTs are still worth investigating: 1) Owing to the device structure and gate technology differences, the 1$^{st}$ BD mechanism varies in the literature as in Ref. [14, 21-23]. Further sweeps and BDs can help to determine the 1$^{st}$ BD mechanism; and 2) To protect a switch system, it is vital to maintain the OFF-state blocking capability (i.e., a low $I_{DSS}$) of the *p*-GaN HEMTs after the gate BD as argued in Ref. [22].

### B. Passivation related BD and PiN BD in the 2$^{nd}$ sweep

As shown in Fig 1(b), after the 2$^{nd}$ BD, $I_G$ increases significantly under both forward and reverse bias. However, either PiN junction BD or passivation related BD may possibly lead to this phenomenon, which makes it challenging to identify the BD mechanism. In this work, we propose a simple but effective method to determine the mechanism, which is to measure the gate-drain current ($I_{GD}$) and the gate-source current ($I_{GS}$) separately. Fig. 3(a) shows $I_G$, $I_{GD}$ and $I_{GS}$ after the 2$^{nd}$ BD in the 3$^{rd}$ sweep. The $I_{GS}$ component is very close to $I_{GD}$ when $V_{GS} = V_{GD} > V_{TH}$ (~ 1.4 V), while $I_{GS}$ is approximately seven orders of magnitude higher than $I_{GD}$ when $V_G < V_{TH}$. This $I_{GD}/I_{GS}$ difference can be explained by the failure of the passivation on the source side. A leakage path between the G and S terminals has been created either along the passivation/*p*-GaN left sidewall or through the passivation in the 2$^{nd}$ BD, while the PiN junction below remains intact. The "gate control" of the channel is still preserved. When $V_G < V_{TH}$, the channel under the gate is depleted; thus, $I_{GD}$ maintains at a low level. When $V_G > V_{TH}$, the channel is turned on connecting the S and D terminals, thus $I_{GD}$ is comparable to $I_{GS}$. This conclusion is confirmed by $I_{DSS}$ in Fig. 3(b). Despite the large reverse $I_G$, the $I_{DSS}$ maintains at a low level after the 2$^{nd}$ BD.

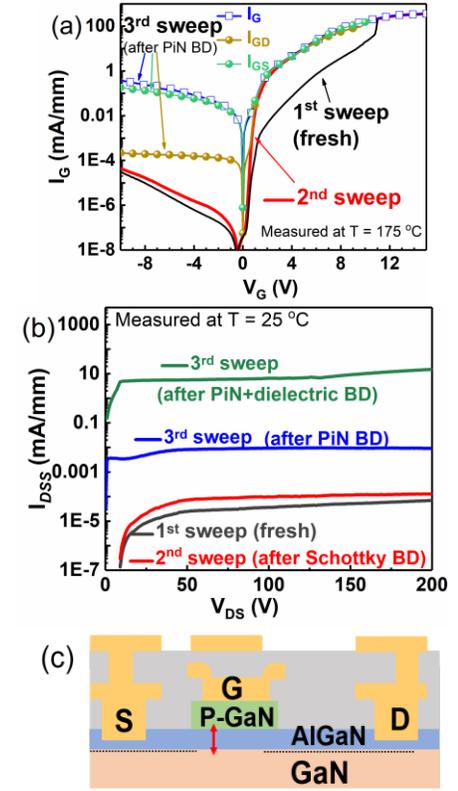

Fig. 4. (a) The PiN junction BD in the 2$^{nd}$ sweep; (b) $I_{DSS}$ before and after PiN junction failure; and (c) illustration of the leakage path responsible for the PiN junction BD.

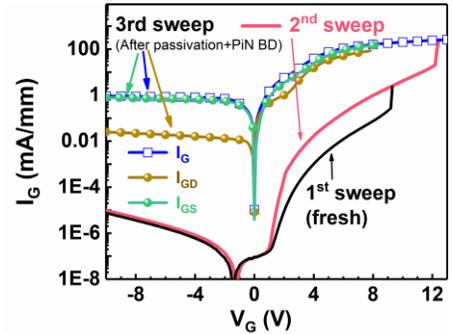

Fig. 5. Gate leakage characteristics before and after the both breakdown of the passivation and PiN junction.

Furthermore, the PiN junction or AlGaN barrier BD has been successfully observed at higher temperatures thanks to the multiple-sweep measurement instead of step stress or a constant stress measurement. As illustrated in Fig. 4(a), in the 2$^{nd}$ sweep at 175 °C (red), there is no typical abrupt $I_G$ increase when $V_{GS} > 0$ V as an indication of a BD. However, in the 3$^{rd}$ sweep, $I_G$ increased several orders of magnitude in the $V_{GS} < 0$ V regime without any $I_G$ increased when $V_{GS} > 0$ V. Besides, both $I_{GD}$ and $I_{GS}$ increased when $V_{GS} < 0$ V. The $I_{DSS}$ also increased several orders of magnitude as shown in Fig. 4(b). Based on these, we can infer that a PiN junction BD happened in the 2$^{nd}$ sweep. Its failure will not contribute more leakage current when $V_{GS} > 0$ V. This phenomenon confirms the possibility of the PiN junction failure and reveals that this failure itself will unlikely result in the increase of $I_G$ in the positive bias regime, disapproving the deductions in previous studies [12-14, 20-22].

This feature makes it difficult to observe the PiN junction BD in a stress measurement due to the lack of a typical breakdown feature. To our best knowledge, this is the first report of direct observation of PiN junction failure. The physical origin PiN BD may be closely related to dislocations in GaN/AlGaN. In literature, it's found that the dislocations in GaN diode can increase the reverse-bias leakage current significantly, whereas they have little impact on the forward-bias current [24]. This conclusion is consistent with our finding that the PiN junction breakdown will only increase the reverse-bias leakage current.

Meanwhile, it's also possible that the passivation BD and the PiN junction BD happen in the same sweep, as illustrated in Fig. 5. The features of different BD mechanisms are summarized in Table I, where the "↑" indicating an increase of the leakage current and the "-" indicating no significant change.

TABLE I
FEATURES OF DIFFERENT BD

|  | when $V_G > 0V$ | | when $V_G < 0V$ | | |
| --- | --- | --- | --- | --- | --- |
|  | $I_{GS}$ | $I_{GD}$ | $I_{GS}$ | $I_{GD}$ | $I_{DSS}$ |
| Schottky BD | ↑ | ↑ | - | - | - |
| Passivation BD* | ↑ | ↑ | ↑ | - | - |
| PiN BD | - | - | ↑ | ↑ | ↑ |
| Passivation+PiN BD | ↑ | ↑ | ↑ | ↑ | ↑ |

When a device is exposed to higher gate voltage, damages are likely to occur due to the heating effect. The cross-sections of the damaged parts have been prepared by focused ion beam etching (FIB) and imaged by scanning electron microscopy (SEM), as illustrated in Fig. 6. Damages originated from the PiN junction failure are clearly shown in Fig. 6(a). Meanwhile, the PiN junction remains intact while the passivation was severely damaged in Fig. 6(b). These observations confirmed our previous deductions that both PiN BD and passivation BD can happen in p-GaN gate HEMTs.

### C. Simulation and gate BV comparisons of the different BD mechanisms

The electric field across the gate region has been simulated numerically by Synopsys' Technology Computer Aided Design (TCAD) software as shown in Figure 7. The results of the simulations showed that the electric field can be quite high in two regions of the device: the p-GaN footing and AlGaN/GaN interface, which are close related to the passivation-related failure and PiN junction failure, respectively.

Temperature-dependent measurements were conducted from 25 °C to 175 °C with 37.5 °C per step to get further insights into the BD mechanisms. For each temperature, at least ten devices were measured by multiple-gate-sweep. Fig. 8(a) shows the statistical summary of the gate BVs. The sequence of the BDs depends on the measurement/operation temperature. Schottky BD has the smallest BV, so it happens before others. At a higher temperature, PiN BD is likely to happen prior to the passivation BD as shown in Fig. 4, which may not be the case for lower temperatures. The different BD mechanisms show different temperature dependence. The Schottky failure has very weak temperature dependence, while the passivation-related failure and PiN failure clearly have positive temperature dependence. The positive temperature dependence of PiN junction failure is ascribed to a dominant role of impact ionization, as discussed in [18].

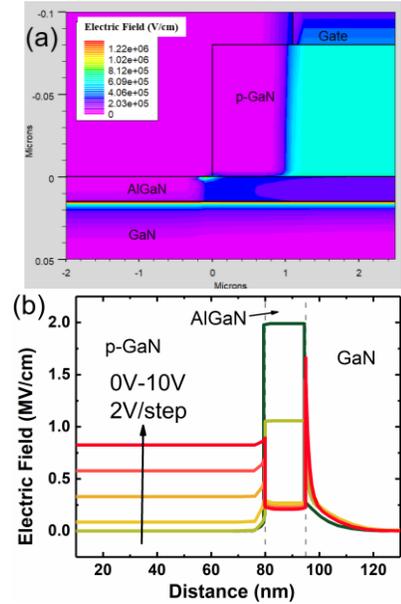

Fig. 7. TCAD Simulation (ATLAS) of the electric field (a) across the gate region under $V_{GS} = 8V$; and (b) across the p-GaN/AlGaN/GaN layers under different gate bias.

Furthermore, the gate BVs with different passivation layers have also been compared. In the comparison group, the device structure and fabrication process are the same as previous except that the $SiN_x$ layer was replaced by a 100 nm $Al_2O_3$ layer deposited by atomic layer deposition (ALD). As illustrated in Fig. 8(b), their Schottky breakdown voltages have little difference at different temperatures, which is as expected since the passivation layer is not related to the Schottky junction. On the other hand, the passivation-related BV in those devices with $Al_2O_3$ passivation layers has a much weaker temperature dependence compared to those with $SiN_x$. These results further valid our BV mechanism determination method discussed above. The "2nd-step" breakdown should be ascribed to passivation related failure instead of the PiN junction failure as claimed by Ref. [21, 22]. Otherwise, the different passivation layers should have little impact on the BVs. These results also help to explain the differences in the temperature dependence of the gate breakdown in literature. Therefore, to improve the gate reliability of p-GaN gate HEMTs, it's also essential to optimize the passivation technology.

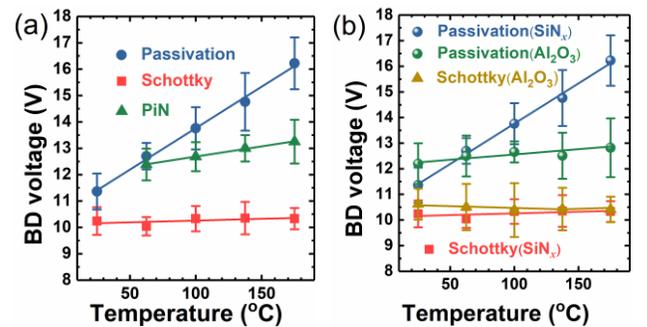

Fig. 8. (a) BVs of different mechanisms and their temperature dependence ($SiN_x$ as the passivation layers); (b) The Schottky junction and passivation related BVs of the devices with a $SiN_x$ or an $Al_2O_3$ as the passivation layer, respectively.

## IV. CONCLUSION

In this work, the gate BD mechanisms of *p*-GaN gate AlGaN/GaN HEMTs were studied thoroughly by the multiple-gate-sweep-based method. Three different BD mechanisms have been observed and confirmed by SEM: the metal/*p*-GaN Schottky junction BD, the PiN junction BD, and the passivation related BD. By measuring $I_{GD}$ and $I_{GS}$ separately and doing $I_{DSS}$ analysis, the BD mechanisms can be successfully identified. We believe this method is generally applicable for p-GaN gate HEMTs. Especially, it's demonstrated that the PiN junction failure alone doesn't lead to an increase of $I_G$ at $V_G > 0$ regime disapproving previous literature. Besides, it's demonstrated that the three BD mechanisms have different temperature dependences. The Schottky junction failure has a very weak temperature dependence, while the PiN junction failure has a positive temperature dependence. For the passivation related breakdown, its dependence on temperature is closely related to the passivation technology. We believe the clarification of BD mechanisms and this BD analysis method will shed more light on improving the gate BV and reliability.